\documentclass[a4]{article}
\usepackage{amssymb,epsfig}

\begin{document}

\title{Comparison of relativistic bound-state calculations in Front-Form and
Instant-Form Dynamics}
\author{B.L.G. Bakker, M. van Iersel, and F. Pijlman \\
	Department of Physics and Astronomy \\
	Vrije Universiteit, Amsterdam, the Netherlands}
%
%\runningauthor{B.L.G. Bakker, M. van Iersel, and F. Pijlman}
%\runningtitle{Comparison of relativistic bound-state calculations in Front-Form
%and Instant-Form Dynamics}
%
%\date{\today}
%
\maketitle
\begin{abstract}
Using the Wick-Cutkosky model and an extended version (massive
exchange) of it, we have calculated the bound states in a quantum field
theoretical approach. In the light-front formalism we have calculated
the bound-state mass spectrum and wave functions. Using the Terent'ev
transformation we can write down an approximation for the angular dependence of
the wave function. After calculating the bound-state spectra we characterized
all states found. Similarly, we have calculated the bound-state
spectrum and wave functions in the instant-form formalism. We compare
the spectra found in both forms of dynamics in the ladder approximation and show
that in both forms of dynamics the $O(4)$ symmetry is broken. 

\end{abstract}
%
%\maketitle
%
\section{Introduction}
\label{sec.01}
Dirac's paper on forms of relativistic dynamics~\cite{Dir49} made it
clear that the real difficulty of constructing a Hamiltonian theory of
interacting particles that satisfies the requirements of special
relativity, is finding the proper form of the interactions. This problem
can be solved in a natural way by resorting to covariant field theory.
Then the construction of all generators of the Poincar\'{e} group can be
found in any textbook on quantum field theory. However, the results
given have usually a formal meaning only and much work needs to be done
to turn them into useful formulas. In particular, drastic
approximations need to be taken, which may or may not be justified. An
approximation that at first sight looks very attractive is to expand
the states of the sytem under consideration into Fock states and
truncate the expansion at a reasonable point. Doing so, one obtains a
quantum-mechanical many-body problem and can use the powerfull machinery
that has been developed for many-body systems.  

The success of Fock-space methods depends crucially on the properties
of the vacuum. Only if one can build on a simple vacuum the Fock-space
expansion is useful. The condition that the vacuum be simple limits
the choice of forms of dynamics in the framework of quantum field
theory essentially to the front form, also known as light-front
dynamics (LFD). In this form three components of the four momentum,
$p^1$, $p^2$, and $p^+ = (p^0 + p^3)/\surd 2$, are independent of the
interaction, while $p^- = (p^0 - p^3)/\surd 2$ contains interaction and
is for this reason said to be a {\em dynamical} operator. The variable
$x^+ = (x^0 + x^3)/\surd 2$ that is conjugate to $p^-$, is the
evolution parameter of states and consequently denoted as light-front
time. The dispersion relation of energy and momentum for a particle
with mass $m$ takes the form $p^- = (\vec{p}^{\,2}_\perp + m^2)/(2 p^+)$,
$\vec{p}_\perp = (p^1, p^2)$. One sees immediately that states of
positive and negative  energy can be separated kinematically, as
positive energy and positive $p^+$ are strictly disconnected. This
property is called the {\em spectrum property}. In instant-form
dynamics (IFD) where $p^0$ is the dynamical component of the four
momentum, the energy may be either positive or negative, independent of
the three momentum. An immediate consequence of the spectrum property is that
massive particles cannot be created from the vacuum in LFD. This is
strictly speaking not enough to reduce the true vacuum to the
perturbative Fock vacuum in LFD and usually, in order to make progress,
one makes the additional assumption that {\em zero modes}, states where
all particles have $p^+ = 0$, are decoupled. In this paper we shall
also make this assumption. For an extensive review of many aspects of
LFD we refer to Brodsky et al.~\cite{BPP98}.

Adopting LFD we still may consider other forms of dynamics, in
particular IFD, for comparison. Even though it is difficult to justify
IFD in a genuine field theoretical setting, to be distinguished from
a situation where binding energies are very much smaller than particle
masses, it remains to be seen what the quantitative differences are
between the results obtained in LFD and IFD respectively. On the level
of calculating invariant amplitudes in a Hamiltonian approach the
situation is clear. One can obtain the time-ordered amplitudes, i.e.,
those that one calculates in LFD or IFD, by integrating Feynman
amplitudes over the relative energy variable, $k^-$ in LFD and $k^0$ in
IFD. Schoonderwoerd et al.~\cite{SBK98} have shown in Yukawa theory
that the difference between the covariant box diagram and the ladder
approximation to it in LFD is much smaller that in IFD. As this result
is obtained for the box diagram with external particles on mass shell
and initial and final state on the energy shell, one cannot immediately
conclude that in a bound-state calculation LFD in the ladder
approximation would be closer to a covariant calculation than IFD.
In this paper we show our results in the simplest possible case, the
Wick-Cutkosky model (WC)~\cite{WC54}. (This model has been reviewed
extensively by Nakanishi~\cite{Nak88}.)

The WC-model is concerned with two scalar fields, $\Phi$ and $\phi$, of
masses $m$ and $\mu$ respectively, with a coupling $g\Phi^* \Phi
\phi$. The original WC-model has $\mu = 0$. We study the bound-state
spectrum and wave functions for a range of coupling-constant values
$g$. Because in LFD, contrary to IFD, the components $L_x$ and $L_y$ of
the orbital angular momentum operator $\vec{L}$ contain interactions,
{\em manifest} rotational invariance is broken in LFD. This means that
if a truncation in Fock space is made, the observables, in particular
the masses of the bound states, will depend on the orientation of the
light front. Below we explain how this violation of rotational
invariance manifests itself in our calculations.

It is well known that the WC-model exhibits $O(4)$ symmetry, like the
nonrelativistic hydrogen atom. This symmetry shows up as a degeneracy
of the levels of different orbital angular momentum $l$. We study the
breaking of this symmetry due to the ladder approximation, which
amounts to the truncation to the two- and three-body sectors in Fock
space. The violation of $O(4)$ symmetry cannot be expected to occur in
LFD only and we show here, for the first time, that it also occurs in
IFD.

The WC-model is popular because it avoids the complications of spin and
an exact solution of the Bethe-Salpeter equation is known in the case
that the mass of the two-body bound state vanishes. Moreover,
Cutkosky~\cite{WC54} gives numerical results for the spectrum in a
range of values of the coupling constant. Recently, Mangin-Brinet et
al.~\cite{MB0001} calculated the ground state of the WC-model in
so-called covariant LFD, also making the ladder approximation. Ji and
Furnstahl~\cite{JF86} estimated the masses of the 1S and 2P states in
the WC-model using a variational Ansatz for the two-body wave function
in LFD, for small values of the coupling constant. Ding and
Darewych~\cite{DD00} also used variational techniques to calculate the
ground state of the WC-model. In the end, they solve an equation  with
a kernel that does not depend on the mass of state, which may explain
the fact that there masses lie much below the masses found either in the
Bethe-Salpeter formalism or in the other Hamiltonian calculations.

Several authors have extended the WC-model by adopting a nonvanishing
mass $\mu$ for the exchanged particle. We mention in particular Cooke
et al.~\cite{CM00}, Nieuwenhuis and Tjon~\cite{NT96}, and again
Refs.~\cite{MB0001,DD00}. In all these works the ratio $\mu/m = 0.15$
close to the ratio of the pion to the proton mass was taken, making
these investigations relevant for the deuteron.

We consider our work as a necessary step to the treatment of more
realistic models, e.g., the Yukawa model. In this domain G{\l}azek et
al.~\cite{G93} have done important pioneering work in LFD. Fuda et al.
 \cite{FUDA} constructed one-boson-exchange models for the
nucleon-nucleon and the pion-nucleon interactions. In the latter case
an LFD calculation was compared to an IFD one. Contrary to what we do,
these authors fit the parameters in the two distinct forms of dynamics
separately to the experimental data, while we keep the model parameters
fixed and compare the spectra.

This paper is organized as follows. In Sec.\ref{sec.02} we give the
details of the model we use and the expressions for the mass operators.
Next we write down the effective two-body equations valid when Fock
space is truncated to the two- and three-particle sectors. In addition
we do not take self-energy terms into account, which would require a
discussion of renormalization, which we want to avoid here. We do not
give a detailed derivation of our equations, as e.g. G{\l}azek et
al.~\cite{G93} have done  so in much detail for the Yukawa model and it
is easy to adapt their methods to the case of scalar particles. In
Sec.\ref{sec.03} we turn to the delicate question how to estimate the
orbital angular momentum of a state in LFD, knowing that rotational
invariance is broken. For the purpose of characterizing the states we rely
on Terent'ev's transformation~\cite{Ter76}, which is known to be exact for
states on the energy shell~\cite{Bak01}. For bound states, which are by
definition not on shell, this transformation can only give tentative results.
The next section contains the details of our numerical methods and in
Sec.\ref{sec.05} we give the masses and the wave functions we found.
Finally we discuss our results and draw conclusions.
\section{Derivation of bound-state equations in the used model}
\label{sec.02}
We use the Wick-Cutkosky model to describe the dynamics of two scalar
particles of opposite charge exchanging a neutral scalar particle. In
the original model by Wick and Cutkosky~\cite{WC54} the exchanged
particle is massless.  It is possible to extend this model to a version
where particles with a certain mass are exchanged. In our model we have
taken bosons of equal mass $m$ and an exchanged particle of mass $\mu$,
which may vanish. The Lagrangian for this system is given by:
\begin{eqnarray}
\mathcal{L} &=& \partial_{\mu}\Phi^{*} \partial^{\mu}\Phi - m^2
\Phi^{*}\Phi + \frac{1}{2} \partial_{\mu} \phi \partial^{\mu} \phi
- \frac{\mu^2}{2} \phi^{2} - g\phi\Phi^{*}\Phi \, ,
\label{eqn1}
\end{eqnarray}
where $g$ is the coupling constant, $\Phi$ the charged field of the
bosons of mass $m$ and $\phi$ the field of the exchanged particle.

In both forms of dynamics we have used a field theoretical method to
derive the bound-state equation from this Lagrangian. In LFD there is
also the possibility to make use of the so-called explicitly covariant
light-front formalism (eg. Ref.~\cite{FUDA, MB0001}). The bound-state
equations in both methods are written in different coordinates, but it
can be shown via a coordinate transformation that the equations are the
same.

Through the energy-momentum tensor it is possible to find an expression
for the Hamiltonian. In LFD the Hamiltonian is given by:
\begin{eqnarray}
 P^{-} &=& \int \!\! [\mathrm{d}^{3} x] \, \left[
 \partial^{\perp} \Phi^{*}\partial^{\perp}\Phi + m^{2}\Phi^{*}\Phi +
 \frac{1}{2}\partial^{\perp}\phi\partial^{\perp}\phi +
 \frac{\mu^2}{2} \phi^{2} + g \phi\Phi^{*} \Phi \right] \, .
\label{eqn2}
\end{eqnarray}
Here $\partial^{\perp} = (\partial^{1},\partial^{2})$ and the integration
element $[\mathrm{d}^{3}x] = \mathrm{d}x^{-} \mathrm{d}^{2} x^{\perp}$.

In IFD the Hamiltonian is given by:
\begin{eqnarray}
H &=& \int \!\! \mathrm{d}^{3} x \left[ \partial^{0} \Phi^{*} \partial^{0}
\Phi + \vec{\nabla} \Phi^{*} \vec{\nabla} \Phi + m^{2} \Phi^{*} \Phi \right. \nonumber\\ 
&+& \left. \frac{1}{2} \partial^{0} \phi \partial^{0} \phi + \frac{1}{2} (\vec{\nabla}
\phi )^{2} + \frac{\mu^{2}}{2} \phi^{2} + g \phi \Phi^{*} \Phi \right]\, .
\label{eqn3}
\end{eqnarray}
Here the integration element $\mathrm{d}^{3}x$ is defined as
$\mathrm{d}^{3}x = \mathrm{d}x^{1}\mathrm{d}x^{2}\mathrm{d}x^{3}$,
as usual.

Following the standard procedures explained in Ref.~\cite{BPP98} for LFD and
Ref.~\cite{itzzub} for IFD, we can expand the free fields $\phi$ and $\Phi$ as
follows
\begin{eqnarray}
\phi(x) \!\!&=&\!\! \int \!\! \{\mathrm{d}^{3}k\}
\left( a(k) e^{-ik x} + a^{\dag}(k) e^{ik x} \right) ,
\label{eqn4} \\
\Phi(x) \!\!&=&\!\! \int \!\! \{\mathrm{d}^{3}k\}
\left( b(k) e^{-ik x} + d^{\dag}(k) e^{ik x} \right) .
\label{eqn5}
\end{eqnarray}
Here $a$ annihilates the exchanged neutral scalar particle, $b$ annihilates the
charged boson and $d$ annihilates the boson with opposite charge.

The difference in the free field expansion in both forms of dynamics lies in
the integration element and in the interpretation of the operators $a$, $b$
and $d$. In LFD the integration element is
\begin{eqnarray}
 \{\mathrm{d}^{3}k\} &=&
 \frac{[\mathrm{d}^{3}k]}{(2\pi)^{3/2}\sqrt{2k^{+}}} \,.
\label{eqn6}
\end{eqnarray}
In LFD we define the integration element in momentum space
$[\mathrm{d}^{3}k]$ in a similar way as we did in coordinate space:
$[\mathrm{d}^{3}k] = \mathrm{d}k^{+} \mathrm{d}^{2}k^{\perp}$. In IFD
the integration element is given by
\begin{eqnarray}
\{\mathrm{d}^{3}k\} &=& \frac{\mathrm{d}^{3}k}{(2\pi)^{3/2}\sqrt{2E(k,m)}} \,.
\label{eqn7}
\end{eqnarray}
Here $\mathrm{d}^{3}k$ is the usual integration element in IFD and the
energy is given by $E(k,m) = \sqrt{\vec{k}^{\,2} + m^{2}}$.

Substituting the free field expansions into Eqs.~(\ref{eqn2}) or (\ref{eqn3}),
we can find an expression for the Hamiltonian. Using normal ordering, we can
write down the Hamiltonian, which can be split into two parts, a free part and a
part containing the interaction. In Eqs.~(\ref{eqn8}) and (\ref{eqn9}) the
expressions are given in LFD:
\begin{eqnarray}
P^{-}_{\mathrm{free}} &=& \int [\mathrm{d}^{3}k] \, \left( \frac{k_{\perp}^{2} + \mu^{2}}
{2k^{+}} a^{\dag}(k)a(k) 
+ \frac{k_{\perp}^{2} + m^{2} }{2k^{+}} \left[b^{\dag}(k)
b(k) + d^{\dag}(k)d(k) \right] \right)\, ,
\label{eqn8} \\
P^{-}_{\mathrm{int}} &=& \frac{g}{(2\pi)^{3/2}} \!\int [\mathrm{d}^{3}k]
[\mathrm{d}^{3} k^{\prime}] \frac{\theta(k^{+}-k^{\prime\,+})}{\sqrt{8k^{+}
k^{\prime\,+}(k^{+}-k^{\prime\,+})}} \nonumber \\
&\times& \hspace{-4mm} \left(a(k-k^{\prime}) \left[ b^{\dag}(k)b(k^{\prime}) +
d^{\dag}(k)d(k^{\prime}) \right] + a^{\dag}(k-k^{\prime}) \left[
b^{\dag}(k^{\prime})b(k) + d^{\dag}(k^{\prime}) d(k) \right] \right) \,.
\label{eqn9}
\end{eqnarray}
The Hamiltonians in IFD are given in Eqs.~(\ref{eqn10}) and (\ref{eqn11}):
\begin{eqnarray}
H_{0} &=& \int \!\! \mathrm{d}^{3}k \left[ E(k,m) \left( b^{\dag}(k)b(k) +
d^{\dag}(k) d(k) \right) + E(k,\mu) a^{\dag}(k)a(k) \right] \, , 
\label{eqn10} \\
H_{\mathrm{int}} &=& \frac{g}{(2\pi)^{3/2}} \int \!\! \mathrm{d}^{3}k
\mathrm{d}^{3}k' \frac{1}{\sqrt{8E(k,m) E(k',m) E(k-k',\mu)}} \nonumber\\
&\times& \hspace{-2mm} \left[ a(k - k') \left( b^{\dag}(k)b(k') + d^{\dag}(k)d(k')
\right) + a^{\dag}(k - k') \left( b^{\dag}(k')b(k) + d^{\dag}(k')d(k)
\right) \right] \, .
\label{eqn11}
\end{eqnarray}
In both forms of dynamics we can write down a Schr\"odinger like
equation. In IFD we have the well-known equation $H|\Psi \rangle =
\mathcal{E}|\Psi \rangle$ where $\mathcal{E}$ is the total energy of the
system and $|\Psi \rangle$ is the total wave function. In LFD the Schr\"odinger
equation is given by
\begin{eqnarray}
\left( 2P^{+}P^{-} - \vec{P}_{\perp}^{2}\right) |\Psi \rangle &=& M^{2}|\Psi \rangle .
\label{eqn12}
\end{eqnarray}
Here $P^{-} = P^{-}_{\mathrm{free}} + P^{-}_{\mathrm{int}}$ is the
total Hamiltonian, $M$ the total mass of the system and $|\Psi \rangle$
is the total wave function.

Since it is not possible to solve a bound-state equation in full Fock
space, we need to make a truncation. Here we follow the same procedure
as G{\l}azek et al.~\cite{G93}. Taking only the two and three particle
sectors into account, we write the total wave function as
\begin{eqnarray}
|\Psi \rangle &=& |\varphi_{2} \rangle + |\varphi_{3} \rangle \, .
\label{eqn13}
\end{eqnarray}
Using this truncation we can write the Schr\"odinger equation as
\begin{eqnarray}
\left( \begin{array}{cc}
 H_{22} & H_{23} \\
 H_{32} & H_{33}
\end{array} \right) \left( \begin{array}{c}
 \varphi_{2} \\ \varphi_{3}
\end{array} \right) &=& E \left( \begin{array}{c}
 \varphi_{2} \\ \varphi_{3}
\end{array} \right) \, .
\label{eqn14}
\end{eqnarray}
Here $H_{22}$, $H_{23}$, $H_{32}$ and $H_{33}$ connect a two particle
state to a two particle state, a three particle state to a two particle
state, etc.

Using this Fock-space truncation and projecting out the states gives two
equations, one for the two-particle sector and one for the three-particle
sector
\begin{eqnarray}
 \langle \varphi_{2}|H_{22}|\varphi_{2} \rangle +  \langle \varphi_{2}|H_{23}|\varphi_{3}
\rangle &=& \langle \varphi_{2}|E|\varphi_{2} \rangle \,,
\label{eqn15} \\
 \langle \varphi_{3}|H_{32}|\varphi_{2} \rangle +  \langle \varphi_{3}|H_{33}|\varphi_{3}
\rangle &=& \langle \varphi_{3}|E|\varphi_{3} \rangle \,.
\label{eqn16}
\end{eqnarray}
Eliminating either the two- or three-particle sector results in a bound-state
equation. We have chosen to eliminate the three-particle sector and write
everything in terms of the two-particle wave function, which results in
\begin{eqnarray}
\left( E - H_{22} \right) |\varphi_{2} \rangle &=& H_{23} \frac{1}{E - H_{33}} H_{32}
\,  |\varphi_{2} \rangle \,.
\label{eqn17}
\end{eqnarray}
The kets $|\varphi_{2} \rangle$ and $|\varphi_{3} \rangle$ in
Eq.~(\ref{eqn13}) can be written in terms of a wave function and
creation operators acting on the vacuum. We have chosen to use two
different bosons in the two- and three-particle states. Then we find richer
spectra, as symmetrization would remove anti-symmetric states. In the
light-front formalism the wave functions are given by
\begin{eqnarray}
|\varphi_{2} \rangle &=& \int \!\! [\mathrm{d}^{3}p]
\varphi_{2}(p,P-p) b^{\dag}(p)d^{\dag}(P-p)|0 \rangle \, ,
\label{eqn18} \\
|\varphi_{3} \rangle &=& \int \!\! [\mathrm{d}^{3}p] [\mathrm{d}^{3}p'] \varphi_{3}
(p,P-p',p'-p) b^{\dag}(p)d^{\dag}(P-p')a^{\dag}(p'-p)|0 \rangle \,.
\label{eqn19}
\end{eqnarray}
Inserting these expressions into the bound-state equation and
truncating after the second order in the coupling constant leads to an
equation with four different contributions to the kernel.  Two of these
can be associated with the self energy and are ignored in this paper.
Only the terms which can be associated with the exchange of a particle
between different constituents are taken into account. This means that
we are working in the ladder approximation and the bound-state equation
in this case becomes
\begin{eqnarray}
\left[ M^{2} - \frac{\vec{p}_{\perp}^{\,2} +m^{2}}{x(1-x)} \right]
\varphi_{2}(\vec{p}_{\perp},x) = \frac{g^{2}}{(2\pi)^{3}} \int\!\! \mathrm{d}^{2}
\vec{p}_{\perp}^{\,\prime}\mathrm{d}x^{\prime}
K(\vec{p}_{\perp},x;\vec{p}_{\perp}^{\,\prime},x^{\prime})
\varphi_{2}(\vec{p}_{\perp}^{\,\prime},x^{\prime}) \, .
\label{eqn20}
\end{eqnarray}
Here $K(\vec{p}_{\perp},x; \vec{p}_{\perp}^{\,\prime},x^{\prime})$ is
the kernel of the equation, $g$ the coupling constant and the relative
variables $x$ and $\vec{p}_{\perp}$ are defined as: $x = p_{1}^{+}/(p_{1}^{+} +
p_{2}^{+})$, $\vec{p}_{\perp} = (1-x)\vec{p}_{1,\perp} -x\vec{p}_{2,\perp}$.
(Note that we used an imaginary coupling to obtain an attractive interaction.)

\begin{figure}
\begin{center}
\epsfig{figure=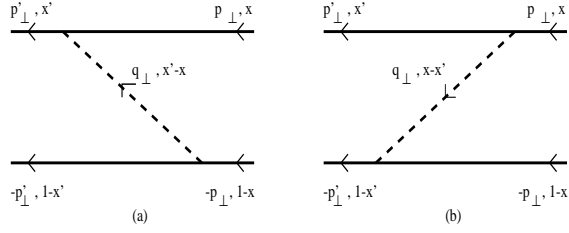,width=7.5cm, height=3cm, angle=0}
\end{center}
\caption{First order time-ordered diagrams in the ladder approximation
for one particle exchange.}
\label{todiag}
\end{figure}

The two terms present in the kernel of the bound-state equation can be
represented graphically by two time-ordered diagrams (Fig.~\ref{todiag}).
The two time-ordered diagrams arise due to the condition that the
plus-momentum of the exchanged particle should be larger than zero
($p^{\prime\,+} - p^{+} > 0$ or $p^{+} - p^{\prime\,+} > 0$). Each of
these diagrams corresponds to an energy denominator which is present in
the kernel of the bound-state equation. The full kernel is given by
\begin{eqnarray}
K(\vec{p}_{\perp},x;\vec{p}_{\perp}^{\,\prime},x^{\prime}) &=&
\frac{1}{\sqrt{x(1-x)x'(1-x')}} \left( \frac{\theta(x'-x)}{2(x'-x)D_{a}} +
\frac{\theta(x-x')}{2(x-x')D_{b}} \right) \, .
\label{eqn21}
\end{eqnarray}
Here the theta-function gives the time-ordering and $D_{a}$ and $D_{b}$
are the energy denominators corresponding to left hand side and right
hand side graph in Fig.~\ref{todiag}. The expressions for these energy
denominators are:
\begin{eqnarray}
 D_{a} &=& M^{2} - \frac{\vec{p}_{\perp}^{\,2}+m^{2}}{x} -
 \frac{\vec{p}_{\perp}^{\,\prime\,2} + m^{2}}{1-x'}
 - \frac{(\vec{p}_{\perp}^{\,\prime} - \vec{p}_{\perp})^{2} + \mu^{2}}{x'-x}
 \, , \label{eqn22} \\
 D_{b} &=& M^{2} - \frac{\vec{p}_{\perp}^{\,\prime\,2}+m^{2}}{x'} -
 \frac{\vec{p}_{\perp}^{\,2} + m^{2}}{1-x}
 - \frac{(\vec{p}_{\perp}-\vec{p}_{\perp}^{\,\prime})^{2}+\mu^{2}}{x-x'}\, .
\label{eqn23}
\end{eqnarray}
In IFD the kets $|\varphi_{2} \rangle$ and $|\varphi_{3} \rangle$ in
Eq.~(\ref{eqn13}) can be written as
\begin{eqnarray}
| \tilde{\varphi}_{2}  \rangle &=& \int \!\! \mathrm{d}^{3} p \tilde{
\varphi}_{2}(p) b^{\dag}(p) d^{\dag}(-p) |0 \rangle \, ,
\label{eqn24} \\
| \tilde{\varphi}_{3}  \rangle &=& \int \!\! \mathrm{d}^{3} p \mathrm{d}^{3} p'
\tilde{\varphi}_{3}(p,p') b^{\dag}(p) d^{\dag}(-p') a^{\dag}(p'-p) |0 \rangle \,.
\label{eqn25}
\end{eqnarray}
After doing calculations similar to those described above we get, when working
in the centre of mass system, the bound-state equation in the IFD formalism
\begin{eqnarray}
\left[ \mathcal{E} - 2 \sqrt{\vec{p}^{\,2} + m^{2}} \right] \tilde{\varphi}_{2}
(\vec{p}) &=& \nonumber\\
&& \hspace{-30mm}\frac{g^{2}}{(2\pi)^{3}} \int \!\! \mathrm{d}^{3} p'
\frac{1}{\mathcal{E} - \sqrt{\vec{p}^{\,\prime 2} + m^{2}} -
\sqrt{\vec{p}^{\,2} + m^{2}} - \sqrt{(\vec{p}^{\,\prime} - \vec{p})^{2}
+ \mu^{2}}} \nonumber\\
&& \hspace{-30mm}\times \frac{\tilde{\varphi}_{2}(\vec{p}^{\,\prime})}{4\sqrt{\vec{p}^{\,\prime 2}
+ m^{2}} \sqrt{\vec{p}^{\,2} + m^{2}} \sqrt{(\vec{p}^{\,\prime} - \vec{p})^{2}
+ \mu^{2}}} \,.
\label{eqn26}
\end{eqnarray}
The bound states in the ladder approximation are found by solving
Eq.~(\ref{eqn20}) and (\ref{eqn26}) respectively.
\section{Characterization of states}
\label{sec.03}
After having found the bound states, we must identify the states and
assign quantum numbers to them. This can be done by using the squared
orbital angular momentum operator $\vec{L}^{\,2}$. In the IFD approach
there is no problem to identify the states. The orbital angular
momentum operator $\vec{L}$ is kinematical and both the orbital angular
momentum quantum number $l$ and the magnetic quantum number $m$ are
good quantum numbers. This in contrast to the LFD approach, where
$\vec{L}$ is dynamical. This leads to the fact that only the helicity
$h$ is a good quantum number and that $l$ is not.

It is possible to approximate the $\vec{L}$-operator in LFD by first
transforming the light-cone variables $(x, \vec{p}_{\perp})$ into the
variables $(p_{z}, \vec{p}_{\perp})$.  Note that this object is not a
true three-dimensional vector, as its components do not transform
properly under all 3D rotations.  The transformation we use was first
introduced by Terent'ev~\cite{Ter76} and is given by
\begin{equation}
 x = \frac{\sqrt{m_{1}^{2}+\vec{p}^{\,2}} + p_{z}}
 {\sqrt{m_{1}^{2}+\vec{p}^{\,2}} + \sqrt{m_{2}^{2}+\vec{p}^{\,2}}} \, .
\label{eqn27}
\end{equation}
Note that this transformation is exact for free particles.
Its inverse is given by
\begin{eqnarray}
p_{z} &=& \left( x - \frac{1}{2} \right) \left[ \frac{\vec{p}_{\perp}^{\,2} +
m_{1}^{2}}{x} + \frac{\vec{p}_{\perp}^{\,2} + m_{2}^{2}}{1-x} \right]^{1/2} -
\frac{m_{1}^{2} - m_{2}^{2}}{2\left[ \frac{\vec{p}_{\perp}^{\,2} + m_{1}^{2}}{x} +
\frac{\vec{p}_{\perp}^{\,2} + m_{2}^{2}}{1-x} \right]^{1/2}} \,.
\label{eqn28}
\end{eqnarray}
After making this transformation, we can approximate the orbital angular
momentum operator $\vec{L}$ by using
\begin{eqnarray}
\vec{L} &=& i \vec{p} \times \vec{\nabla}_{p} \,.
\label{eqn29}
\end{eqnarray}
To characterize a state we have found, we calculate the percentage of a
specific angular momentum state (i.e. S, P, D, ...) present in this
state. This is done by taking the inner product with a spherical
harmonic $Y_{lm}$, thus projecting the wave function on a radial
function. In this way we can calculate for every angular momentum
quantum number $l$ whether it is present in a calculated state and if
so, how big its contribution to that state is. If the overlap is over
85\% we characterize the calculated state with the quantum number $l$.
We realize that this way of determining the angular momentum is only
approximate.
\section{Method of solution}
\label{sec.04}
We solved both bound-state equations (in LFD and IFD) numerically. This
was done by integrating out the angular dependence and making an
expansion into basis functions.

The LFD wave function $\varphi_{2}(\vec{p}_{\perp},x)$ in the
bound-state equation, Eq.~(\ref{eqn20}), depends on both the momentum
and the orientation of $\vec{p}_{\perp}$ in space. It is possible to
expand this wave function in eigenfunctions of $L_{z}$,
\begin{eqnarray}
\varphi_{2}(\vec{p}_{\perp},x) &=& \sum_{h} \frac{1}
{\sqrt{\frac{p_{\perp}^{2} + m^{2}}{x(1-x)} - M^{2}}}
\psi_{h}(p_{\perp},x) \chi_{h}(\phi) \,.
\label{eqn30}
\end{eqnarray}
where $h$ is the helicity and the factor $1/\sqrt{\frac{p_{\perp}^{2} +
m^{2}}{x(1-x)} - M^{2}}$ is introduced to symmetrize the integral equation. The
angular part of a wave function with fixed helicity $h$ is
\begin{eqnarray}
 \chi_{h}(\phi) &=& \frac{1}{\sqrt{2\pi}} \exp(-ih\phi) \,.
\label{eqn31}
\end{eqnarray}
We use it to integrate out the angular dependence in the integral
equation, Eq.~(\ref{eqn20}).

As we can see from Eq.~(\ref{eqn30}) the momentum dependent wave
functions in LFD depend on both the perpendicular and plus component of
the momenta. For the $x$-dependence of this wave function we have used
cubic spline functions Ref.~\cite{splines, splines2} and for the
$p_{\perp}$-dependence we have used a basis which contains Jacobi
polynomials, viz
\begin{eqnarray}
\psi_{kl}(p_{\perp}) &=& \bar{N}_{kl} p^{l} \left(
\frac{\mu_{sc}}{p^{2}+\mu^{2}_{sc}} \right)^{l+3/2} P_{k}^{(l+1,l)}
\left(\frac{p^{2}-\mu^{2}_{sc}}{p^{2}+\mu^{2}_{sc}} \right) \, .
\label{eqn32}
\end{eqnarray}
Here $\mu_{sc}$ is a scaling parameter, $l = 0, 1, 2, ...$, and
$\bar{N}_{kl}$ is the normalization constant, which is given by
\begin{eqnarray}
\bar{N}_{kl} &=& \frac{2\sqrt{\mu_{sc} k! (k+2l+1)!}}{\Gamma(k+l+1)} \, .
\label{eqn33}
\end{eqnarray}
These functions are similar to functions which were first introduced by
Olsson and Weniger (see below). We have chosen them because they have
the right behaviour at the origin and for $p \rightarrow \infty$.

For technical reasons we symmetrized the integral equation using the
transformation given in Eq.~(\ref{eqn30}). The kernel of the symmetrized
bound-state equation for fixed helicity $h$ is given by:
\begin{eqnarray}
 K(p_{\perp},x;p_{\perp}^{\,\prime},x') &=& \frac{p_{\perp}^{3/2}p_{\perp}^{\prime\,3/2}}
 {\sqrt{M^{2}x(1-x) - (p_{\perp}+m^{2})}\sqrt{M^{2}x'(1-x') - (p_{\perp}^{\prime\,2}
 + m^{2})}} \nonumber\\
 &&\hspace{-30mm}\times \left[ \frac{2\pi\theta(x'-x)}{2(x'-x)D'_{a}} \frac{1}{\sqrt{1 - \left(
 \frac{2p_{\perp}p_{\perp}'}{(x'-x)D'_{a}}\right)^{2}}} \left( \frac{\sqrt{1 - \left(
 \frac{2p_{\perp}p_{\perp}'}{(x'-x)D'_{a}} \right)^{2}} - 1}
 {\frac{2p_{\perp}p_{\perp}'}{(x'-x)D'_{a}}} \right)^{|h|} \right. \nonumber\\
 && \left. \hspace{-30mm} + \frac{2\pi\theta(x-x')}{2(x-x')D'_{b}} \frac{1}{\sqrt{1 - \left(
 \frac{2p_{\perp}p_{\perp}'}{(x-x')D'_{b}} \right)^{2}}} \left( \frac{\sqrt{1 - \left(
 \frac{2p_{\perp}p_{\perp}'}{(x-x')D'_{b}} \right)^{2}} - 1}
 {\frac{2p_{\perp}p_{\perp}'}{(x-x')D'_{b}}} \right)^{|h|} \right]\,,
 \label{eqn34}
\end{eqnarray}
where $D'_{a}$ and $D'_{b}$ are given by
\begin{eqnarray}
D'_{a} &=& M^{2} - \frac{p_{\perp}^{2}+m^{2}}{x} -
\frac{p_{\perp}^{\prime\,2} + m^{2}}{1-x'}
- \frac{p_{\perp}^{\prime\,2} + p_{\perp}^{2} + \mu^{2}}{x'-x}
\, , \label{eqn35} \\
D'_{b} &=& M^{2} - \frac{p_{\perp}^{\prime\,2}+m^{2}}{x'} -
\frac{p_{\perp}^{2} + m^{2}}{1-x}
- \frac{p_{\perp}^{2} + p_{\perp}^{\prime\,2} + \mu^{2}}{x-x'}\, .
\label{eqn36}
\end{eqnarray}
Since we have integrated out the angular dependence the vector
character of the momenta in these equations has disappeared. Note that
in the kernel the absolute value of the helicity is present, which
implies that states with opposite helicities are degenerate. 

The wave functions in IFD depend on the three-momentum (see e.g.
Eq.~(\ref{eqn26})). In this case we can expand the wave functions into
spherical harmonics
\begin{eqnarray}
\tilde{\varphi}_{2}(\vec{p}) &=& \sum_{lm} Y_{lm}(\hat{p}) \phi_{l}(p) \,.
\label{eqn37}
\end{eqnarray}
Note that we can integrate out both angles $\theta$ and $\phi$, whereas
in LFD we could only integrate out one angle, $\phi$. For the momentum
dependent part of the wave function we can make an expansion into basis
functions as well. Here we use a basis which was first introduced by
Olsson and Weniger~\cite{olswen}
\begin{eqnarray}
\psi^{OW}(p) &=& N_{kl} p^{l} \left(
\frac{\mu_{sc}}{p^{2}+\mu^{2}_{sc}} \right)^{l+2} P_{k}^{(l+3/2,l+1/2)}
\left(\frac{p^{2}-\mu^{2}_{sc}}{p^{2}+\mu^{2}_{sc}} \right) \, .
\label{eqn38}
\end{eqnarray}
Here $\mu_{sc}$ is a scaling parameter, $l = 0, 1, 2, ...$, and the
normalization constant $N_{kl}$ is given by
\begin{eqnarray}
N_{kl} &=& \frac{2\sqrt{\mu_{sc} k! (k+2l+2)!}}{\Gamma(k+l+3/2)} \, .
\label{eqn39}
\end{eqnarray}
We symmetrize the integral equation by using the transformation
\begin{eqnarray}
\phi_{l}(p) &=& \frac{\bar{\phi}_{l}(p)}{\sqrt{\mathcal{E} - 2\sqrt{p^{2} +
m^{2}}}} \,.
\label{eqn40}
\end{eqnarray}
After symmetrizing the equation and integrating out the angular
dependence the kernel of the bound-state equation in IFD becomes
\begin{eqnarray}
\tilde{K}(p,p')\! &=& \! \frac{2\pi}{\sqrt{\mathcal{E} - 2
\sqrt{p^{2}+m^{2}}}\sqrt{\mathcal{E} - 2\sqrt{p^{\prime\,2}+m^{2}}}}
\frac{V_{l}(p,p')}{4\sqrt{p^{2}+m^{2}}\sqrt{p^{\prime\,2}+m^{2}}} \,.
\label{eqn41}
\end{eqnarray}
Here $l$ is the angular momentum quantum number and $V_{l}(p,p')$ is
obtained by integrating over the angle between $\vec{p}$ and
$\vec{p}^{\,\prime}$ and is given by
\begin{eqnarray}
V_{l}(p,p') &=& \int \frac{\mathrm{d}(\hat{p}\cdot \hat{p'}) P_{l}
(\hat{p}\cdot \hat{p'})}{\mathcal{E} - \sqrt{p^{2}+m^{2}} -
\sqrt{p^{\prime\,2}+m^{2}} - \sqrt{p^{2} + p^{\prime\,2} + \mu^{2}
- 2 p p'(\hat{p}\cdot\hat{p'})}} \nonumber\\
&& \times \frac{1}{\sqrt{p^{2} + p^{\prime\,2}
+ \mu^{2} - 2pp'(\hat{p}\cdot\hat{p'})}} \,,
\label{eqn42}
\end{eqnarray}
where $P_{l}(\hat{p}\cdot\hat{p'})$ is a Legendre polynomial.\\
\section{Numerical results}
\label{sec.05}
First we studied the accuracy of the matrix elements. Using enough
integration points allows us to calculate the matrix elements with an
accuracy of at least six decimal places. We studied the convergence of
the spectra with respect to the number of basis functions, i.e., spline
functions and Olsson Weniger like functions in LFD and the functions
introduced by Olsson and Weniger in the case of IFD. In LFD we found
that 14 spline functions or more and eight Olsson Weniger like functions or
more give an accuracy of at least three decimal places for the lowest state
for any $l$. The states with principal quantum number $n = 3$ have an
estimated absolute error of 0.004. In IFD taking 20 basis functions or
more gives an accuracy of four decimal places in all cases considered.

Taking more basis functions into account gives some variation in the
absolute values of the masses. The relative position of the bound
states does not change, which means that the spectra are not changed
qualitatively.

We have also compared our calculations in LFD with those of
Mangin-Brinet et al.~\cite{MB0001}.  They have calculated the ground-state
masses only and we have compared our calculated ground-state
masses with those given in Ref.~\cite{MB0001}. We have to remark that
in that paper a coupling constant $\alpha$ is used, where $\alpha =
g^{2} / (16 \pi m^{2})$. In Table~\ref{compare} our calculated ground-state
masses are compared with the values from Ref.~\cite{MB0001} for
different values of the coupling constant.
\begin{table}[ht]
\caption{Comparison of the ground-state masses given in
Ref.~\cite{MB0001} (M.B.) and those obtained in the present calculation
for different values of the coupling constant.}
\begin{center}
\begin{tabular}{|c|c|c|c|}
\hline
$\alpha$ & $g$ & mass M.B. & mass this calc. \\
\hline
6.0 & 17.36 & 0.820 & 0.811 \\
5.0 & 15.85 & 1.160 & 1.156 \\
4.0 & 14.18 & 1.415 & 1.412 \\
2.0 & 10.03 & 1.790 & 1.788 \\
1.0 &  7.09 & 1.923 & 1.922 \\
0.5 &  5.01 & 1.973 & 1.973 \\
\hline
\end{tabular}
\end{center}
%tab.01
\label{compare}
\end{table}
For smaller values of $g$ we find the same bound-state masses, while for
larger values we find somewhat smaller masses. As in Ref.~\cite{MB0001}
the masses are given in three decimal places only, we conclude that our
calculations are essentially in agreement with those of Mangin-Brinet
et al.
\subsection{Spectra}
For different values of the coupling constant we have calculated the
spectra in the case of massless exchange. In the LFD formalism we have
calculated the bound states for different helicities and afterwards
determined the corresponding orbital angular momentum quantum number.
Here we have used eight basis functions for the $p_{\perp}$-dependence
and 14 spline functions for the plus-momentum dependence of the
wave function. In the IFD formalism the bound states are calculated
with $l$ as a good quantum number and by using 25 Olsson Weniger
functions.

\begin{figure}
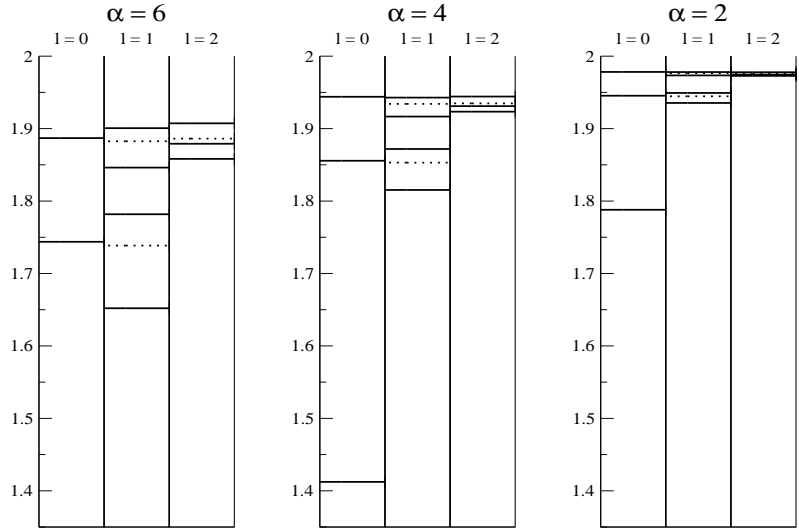

\epsfig{figure=lfdmu0alfa6.eps, width=3cm, height=7cm, angle=0}
\hspace{5mm}
\epsfig{figure=lfdmu0alfa4.eps, width=3cm, height=7cm, angle=0}
\hspace{5mm}
\epsfig{figure=lfdmu0alfa2.eps, width=3cm, height=7cm, angle=0}
\hspace{5mm}
\caption{Spectra in LFD in the case $\mu = 0$ for different coupling
constants.}
\label{speclfd}
\end{figure}

In Fig.~\ref{speclfd} the spectra are plotted for coupling constant $g
= 17.36$, 14.18, and 10.03, which correspond to $\alpha = 6$, $4$, and
$2$. We have only plotted the cases $l = 0, 1, 2$. We should remark
that in Fig.~\ref{speclfd} the spectrum for $\alpha = 6$ is not
completely shown. (We have left out the ground state to get a better
view of the rest of the spectrum.)
\begin{figure}
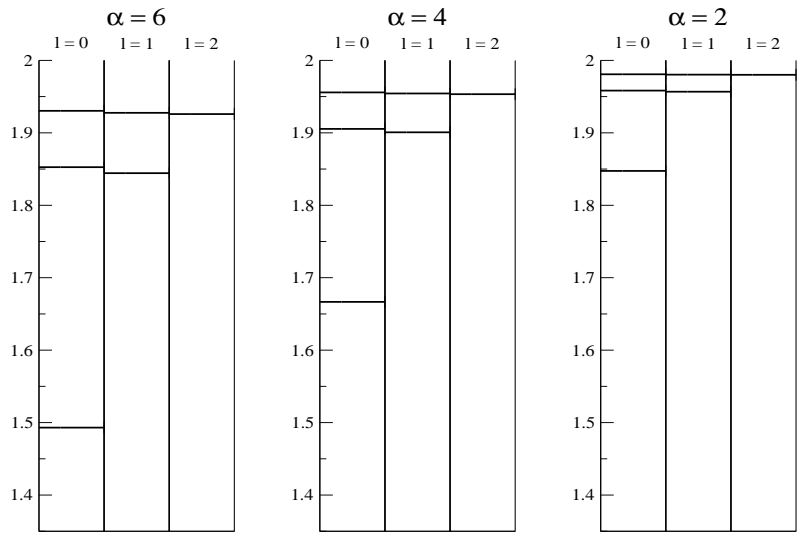

\epsfig{figure=ifdmu0alfa6.eps, width=3cm, height=7cm, angle=0}
\hspace{5mm}
\epsfig{figure=ifdmu0alfa4.eps, width=3cm, height=7cm, angle=0}
\hspace{5mm}
\epsfig{figure=ifdmu0alfa2.eps, width=3cm, height=7cm, angle=0}
\caption{Spectra in IFD in the case $\mu = 0$ for different coupling
constants.}
\label{specifd}
\end{figure}
Due to the breaking of manifest rotational invariance in LFD the
P-states and D-states are split. The center of gravity of these split
levels is plotted in Fig.~\ref{speclfd} as a dashed line and the values
corresponding to this center of gravity are given in
Table~\ref{average}. Looking at the S-states and the dashed lines in
Fig.~\ref{speclfd} we observe a more or less similar pattern as found
in IFD (Fig.~\ref{specifd}).
\begin{table}[ht]
\caption{Masses corresponding to the center of gravity for the P-states and
D-states in LFD for $\mu = 0$ and coupling constants $g = 17.36, 14.18$, and 10.03.}
\begin{center}
\begin{tabular}{|c|c|c|c|c|c|c|c|}
\hline
\multicolumn{2}{|c|}{$g = 17.36$} && \multicolumn{2}{c|}{$g = 14.18$} &&
\multicolumn{2}{c|}{$g = 10.03$} \\
\hline
\multicolumn{2}{|c|}{$l = 0$} && \multicolumn{2}{|c|}{$l = 0$} &&
\multicolumn{2}{|c|}{$l = 0$} \\
\hline
0.811 & 1S && 1.412 & 1S && 1.788 & 1S \\
1.744 & 2S && 1.856 & 2S && 1.945 & 2S \\
1.885 & 3S && 1.936 & 3S && 1.978 & 3S \\
\hline
\multicolumn{2}{|c|}{$l = 1$} && \multicolumn{2}{|c|}{$l = 1$} &&
\multicolumn{2}{|c|}{$l = 1$} \\
\hline
1.739 & 2P && 1.853 & 2P && 1.945 & 2P \\
1.883 & 3P && 1.934 & 3P && 1.977 & 3P \\
\hline
\multicolumn{2}{|c|}{$l = 2$} && \multicolumn{2}{|c|}{$l = 2$} &&
\multicolumn{2}{|c|}{$l = 2$} \\
\hline
1.886 & 3D && 1.935 & 3D && 1.976 & 3D \\
\hline
\end{tabular}
\end{center}
%tab.02
\label{average}
\end{table}
Looking at the spectra for different coupling constants we see that the
binding becomes less when the coupling constant becomes smaller. This
is the same in both IFD and LFD. Comparing between the different
formalisms we see that the states in the LFD formalism have more
binding.

If instead of massless exchange $\mu \neq 0$ is considered, one may
expect the masses to increase. This is borne out by comparing $\mu = 0$
and $\mu = 0.15$. The comparison is done for all three values of the
coupling constant used above, i.e., $\alpha = 6$, 4 and 2. In
Table~\ref{alfa6} the LFD calculated masses with orbital angular
momentum quantum number $l = 0, 1, 2$ ($\mu = 0$ and $\mu = 0.15$) are
compared to those calculated in IFD for a coupling constant $g = 17.36$
($\alpha = 6$). The calculated masses for a coupling constant $g =
14.18$ ($\alpha = 4$) and $g = 10.03$ ($\alpha = 2$) are given in
Table~\ref{alfa4} and \ref{alfa2}.  In all these tables the quantum
numbers of the states are given as well as the masses.

\begin{table}[ht]
\caption{Comparison of the calculated mass spectra in LFD and IFD in the case of
massless and massive exchange for a coupling constant $g = 17.36$ ($\alpha
= 6$).}
\begin{center}
\begin{tabular}{|c|c|c|c|c|c|c|c|c|c|c|c|}
\hline
\multicolumn{12}{|c|}{$\mu = 0.0$}\\
\hline
\multicolumn{2}{|c|}{LFD} & \multicolumn{2}{c|}{IFD} & \multicolumn{2}{c|}{LFD} &
\multicolumn{2}{c|}{IFD} & \multicolumn{2}{c|}{LFD} & \multicolumn{2}{c|}{IFD} \\
\hline
\multicolumn{4}{|c|}{$l = 0$} & \multicolumn{4}{c|}{$l = 1$} & \multicolumn{4}{c|}{$l = 2$} \\
\hline
0.811 & 1S & 1.4931 & 1S & 1.652 & 2P & 1.8443 & 2P & 1.858 & 3D & 1.9259 & 3D \\
1.744 & 2S & 1.8526 & 2S & 1.782 & 2P &        &    & 1.879 & 3D &        &    \\
1.885 & 3S & 1.9304 & 3S & 1.846 & 3P & 1.9276 & 3P & 1.907 & 3D &        &    \\
   -  &    &    -   &    & 1.901 & 3P &        &    &    -  &    &    -   &    \\
\hline
\hline
\multicolumn{12}{|c|}{$\mu = 0.15$}\\
\hline
\multicolumn{2}{|c|}{LFD} & \multicolumn{2}{c|}{IFD} & \multicolumn{2}{c|}{LFD} &
\multicolumn{2}{c|}{IFD} & \multicolumn{2}{c|}{LFD} & \multicolumn{2}{c|}{IFD} \\
\hline
\multicolumn{4}{|c|}{$l = 0$} & \multicolumn{4}{c|}{$l = 1$} & \multicolumn{4}{c|}{$l = 2$} \\
\hline
0.923 & 1S & 1.5384 & 1S & 1.753 & 2P & 1.9153 & 2P &  - &    & - & \\
1.846 & 2S & 1.9276 & 2S & 1.870 & 2P &        &    &  - &    & - & \\
\hline
\end{tabular}
\end{center}
%tab.03
\label{alfa6}
\end{table}
\begin{table}[ht]
\caption{Comparison of the calculated bound-state masses in LFD and IFD in the
case of massless and massive exchange for a coupling constant $g = 14.18$
($\alpha = 4$).}
\begin{center}
\begin{tabular}{|c|c|c|c|c|c|c|c|c|c|c|c|}
\hline
\multicolumn{12}{|c|}{$\mu = 0.0$}\\
\hline
\multicolumn{2}{|c|}{LFD} & \multicolumn{2}{c|}{IFD} & \multicolumn{2}{c|}{LFD} &
\multicolumn{2}{c|}{IFD} & \multicolumn{2}{c|}{LFD} & \multicolumn{2}{c|}{IFD} \\
\hline
\multicolumn{4}{|c|}{$l = 0$} & \multicolumn{4}{c|}{$l = 1$} & \multicolumn{4}{c|}{$l = 2$} \\
\hline
1.412 & 1S & 1.6666 & 1S & 1.815 & 2P & 1.9007 & 2P & 1.923 & 3D & 1.9534 & 3D \\
1.856 & 2S & 1.9054 & 2S & 1.872 & 2P &        &    & 1.931 & 3D &        & \\
1.944 & 3S & 1.9558 & 3S & 1.917 & 3P & 1.9543 & 3P & 1.944 & 3D &        & \\
   -  &    &    -   &    & 1.943 & 3P &        &    &    -  &    &    -   & \\
\hline
\hline
\multicolumn{12}{|c|}{$\mu = 0.15$}\\
\hline
\multicolumn{2}{|c|}{LFD} & \multicolumn{2}{c|}{IFD} & \multicolumn{2}{c|}{LFD} &
\multicolumn{2}{c|}{IFD} & \multicolumn{2}{c|}{LFD} & \multicolumn{2}{c|}{IFD} \\
\hline
\multicolumn{4}{|c|}{$l = 0$} & \multicolumn{4}{c|}{$l = 1$} & \multicolumn{4}{c|}{$l = 2$} \\
\hline
1.477 & 1S & 1.7093 & 1S & 1.898 & 2P & 1.9628 & 2P & - & & - & \\
1.937 & 2S & 1.9690 & 2S & 1.945 & 2P &        &    & - & & - & \\
\hline
\end{tabular}
\end{center}
%tab.04
\label{alfa4}
\end{table}
\begin{table}[ht]
\caption{Comparison of the calculated bound-state masses in LFD and IFD in the
case of massless and massive exchange for a coupling constant $g = 10.03$
($\alpha = 2$).}
\begin{center}
\begin{tabular}{|c|c|c|c|c|c|c|c|c|c|c|c|}
\hline
\multicolumn{12}{|c|}{$\mu = 0.0$}\\
\hline
\multicolumn{2}{|c|}{LFD} & \multicolumn{2}{c|}{IFD} & \multicolumn{2}{c|}{LFD} &
\multicolumn{2}{c|}{IFD} & \multicolumn{2}{c|}{LFD} & \multicolumn{2}{c|}{IFD} \\
\hline
\multicolumn{4}{|c|}{$l = 0$} & \multicolumn{4}{c|}{$l = 1$} & \multicolumn{4}{c|}{$l = 2$} \\
\hline
1.788 & 1S & 1.8474 & 1S & 1.935 & 2P & 1.9568 & 2P & 1.973 & 3D & 1.9801 & 3D \\
1.945 & 2S & 1.9583 & 2S & 1.949 & 2P &        &    & 1.975 & 3D &        & \\
1.978 & 3S & 1.9808 & 3S & 1.973 & 3P & 1.9803 & 3P & 1.978 & 3D &        & \\
   -  &    &    -   &    & 1.978 & 3P &        &    &    -  &    &    -   & \\
\hline
\hline
\multicolumn{12}{|c|}{$\mu = 0.15$}\\
\hline
\multicolumn{2}{|c|}{LFD} & \multicolumn{2}{c|}{IFD} & \multicolumn{2}{c|}{LFD} &
\multicolumn{2}{c|}{IFD} & \multicolumn{2}{c|}{LFD} & \multicolumn{2}{c|}{IFD} \\
\hline
\multicolumn{4}{|c|}{$l = 0$} & \multicolumn{4}{c|}{$l = 1$} & \multicolumn{4}{c|}{$l = 2$} \\
\hline
1.833 & 1S & 1.8849 & 1S & 1.992 & 2P & 1.9995 & 2P & - & & - & \\
1.995 & 2S & 1.9982 & 2S & 1.998 & 2P &   -    &    & - & & - & \\
\hline
\end{tabular}
\end{center}
%tab.05
\label{alfa2}
\end{table}
As we already saw in Fig.~\ref{speclfd}, the states have less binding
for decreasing coupling constant and we find only a few bound states in
the case of $g = 10.03$. Tables~\ref{alfa6}, \ref{alfa4}, and
\ref{alfa2} show the same effect. From these results it also becomes
clear that the states in the case of massless exchange have more
binding than those for $\mu = 0.15$. Especially in the case of $l = 1$
or 2 the binding is sometimes so small that we did not succeed in
calculating the bound-state mass with the number of basis functions we have used.
\subsection{Wave functions}
\begin{figure}[h]
\epsfig{figure=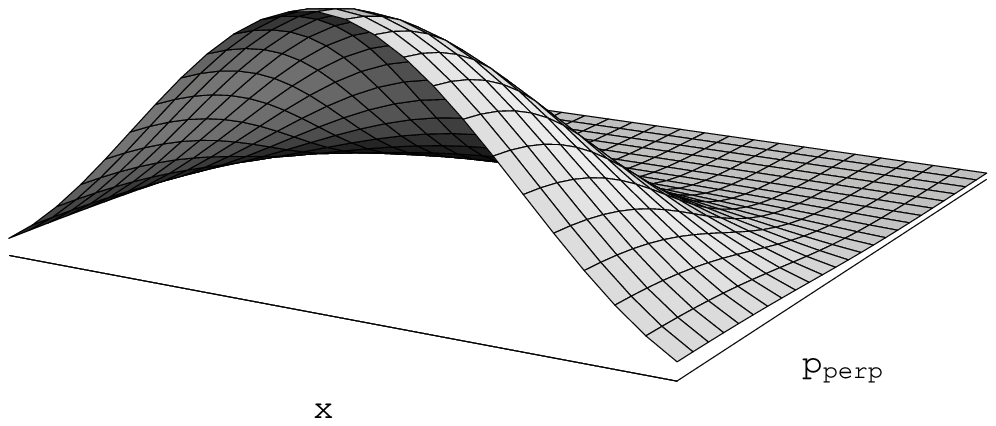, width=4cm, height=4cm, angle=0}
\hspace{5mm}
\epsfig{figure=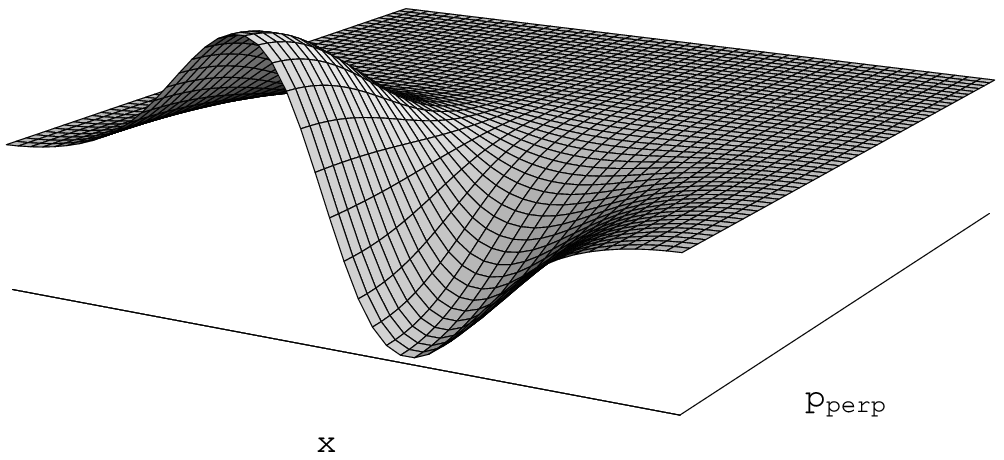, width=4cm, height=4cm, angle=0}
\hspace{5mm}
\epsfig{figure=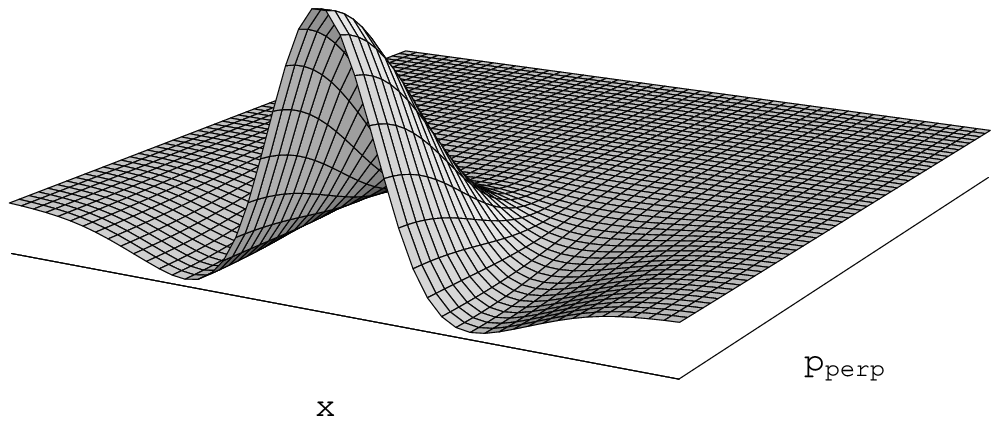, width=4cm, height=4cm, angle=0}
\caption{LFD wave functions (1S, 2P, 2S) corresponding to the first three states
in the case $m = 1.0$, $\mu = 0.0$ and $g = 17.36$}
\label{3dwaves}
\end{figure}

Besides the spectra we calculated the wave functions corresponding to
the bound states. In Fig.~\ref{3dwaves} the 2-D LFD wave functions of
the 1S, 2S and 2P states for $g = 17.36$ and $\mu = 0$ are given.
These figures show the variation of the wave function with $x$ (or
$p^{+}$) and $|\vec{p}_{\perp}|$. We see that the wave function goes
asymptotically to zero when $|\vec{p}_{\perp}|$ becomes large.

\begin{figure}[h]
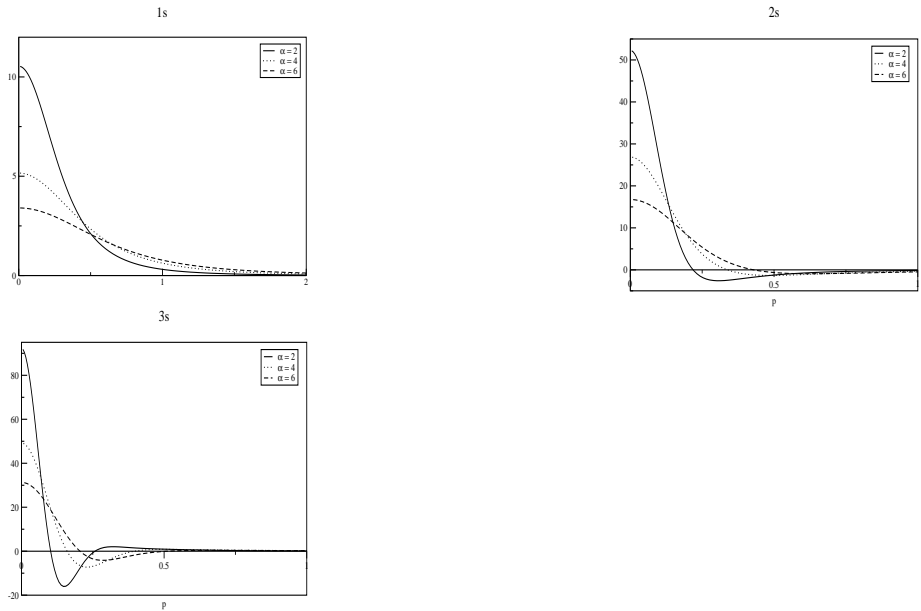

\epsfig{figure=all-1s.eps, width=4cm, height=4cm, angle=0}
\hspace{5mm}
\epsfig{figure=all-2s.eps, width=4cm, height=4cm, angle=0}
\hspace{5mm}
\epsfig{figure=all-3s.eps, width=4cm, height=4cm, angle=0}
\caption{LFD radial wave functions corresponding to the 1S, 2S and 3S
states for $\alpha = 6$, 4 and 2.}
\label{lfdsradial}
\end{figure}

In Fig.~\ref{lfdsradial} the wave functions corresponding to the 1S, 2S,
and 3S states are plotted for all three values of the coupling constant
in the case of massless exchange.
This figure shows that the wave function becomes flatter and broader
when the coupling constant increases. All functions go asymptotically
to zero and show the correct number of nodes; i.e. zero nodes for the
1S state, one node for the 2S-state etc.

\begin{figure}[h]
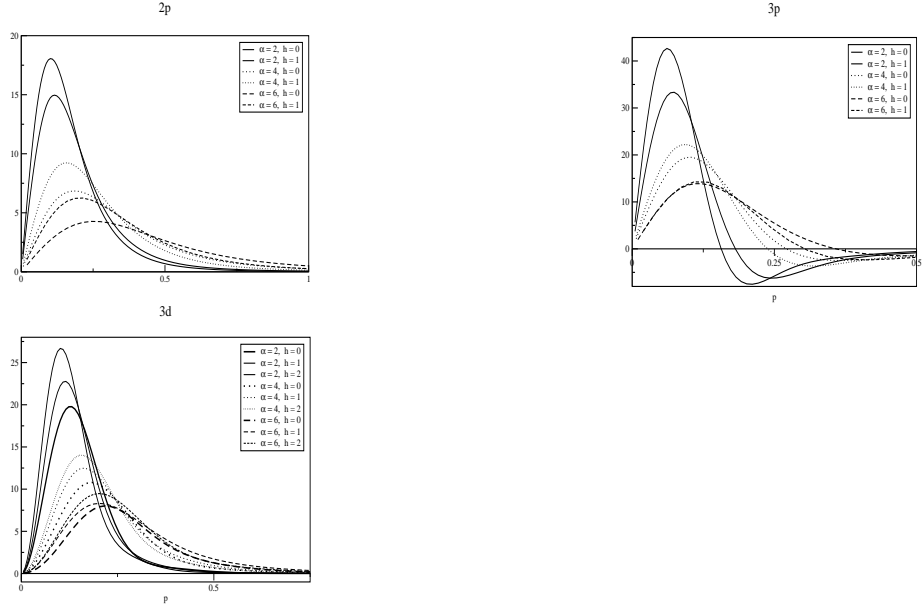

\epsfig{figure=all-2p.eps, width=4cm, height=4cm, angle=0}
\hspace{5mm}
\epsfig{figure=all-3p.eps, width=4cm, height=4cm, angle=0}
\hspace{5mm}
\epsfig{figure=all-3d.eps, width=4cm, height=4cm, angle=0}
\caption{LFD radial wave functions corresponding to the 2P, 3P and
3D states for $\alpha = 6$, 4 and 2.}
\label{lfdradial}
\end{figure}

\begin{figure}[h]
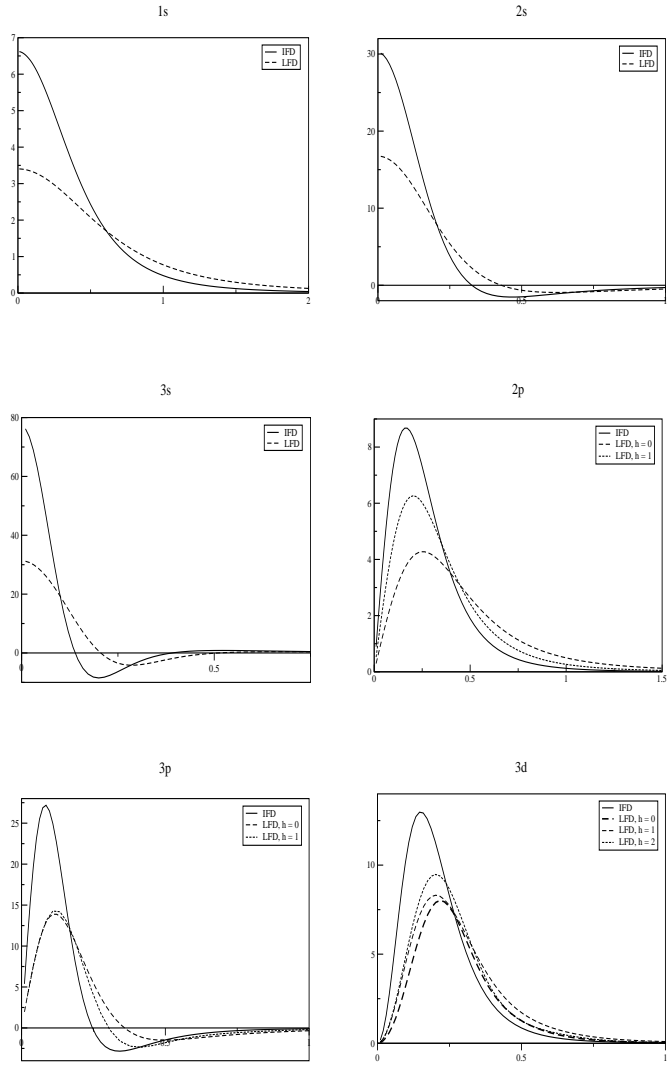

\begin{center}
\epsfig{figure=comp1s.eps, width=4cm, height=4cm, angle=0}
\hspace{5mm}
\epsfig{figure=comp2s.eps, width=4cm, height=4cm, angle=0}

\vspace{10mm}
\epsfig{figure=comp3s.eps, width=4cm, height=4cm, angle=0}
\hspace{5mm}
\epsfig{figure=comp2p.eps, width=4cm, height=4cm, angle=0}

\vspace{10mm}
\epsfig{figure=comp3p.eps, width=4cm, height=4cm, angle=0}
\hspace{5mm}
\epsfig{figure=comp3d.eps, width=4cm, height=4cm, angle=0}

\vspace{10mm}
\caption{Comparison of the radial wave functions calculated in LFD and
IFD at a coupling constant $g = 17.36$.}
\end{center}
\label{radcomp}
\end{figure}

In the case of the S-states, the helicity can only be equal to zero.
This in contrast to the P- and D-states where we have helicity $-1$, 0
and 1 (P-states) and $-2$, $-1$, 0, 1 and 2 (D-states). In our
calculations we have worked with positive helicities only. We can do
this because of the fact that only the absolute value of the helicity
is present in the kernel, Eq.~(\ref{eqn34}). In Fig.~\ref{lfdradial}
the 2P, 3P, and 3D states for all (positive) helicities are plotted in
the case of $\mu = 0$ for coupling constant $\alpha = 6, 4, 2$. As
could be expected, the radial wave functions depend on the helicity,
which is again due to the breaking of rotational invariance.
The comparison between the radial wave functions calculated in LFD and
IFD is made in Fig.~\ref{radcomp}. In this figure the wave functions
are plotted for a coupling constant $g = 17.36$ only.
Comparing the plots in Fig.~\ref{radcomp} we see that the wave function
calculated in the LFD formalism is flatter and broader than the wave
function calculated in IFD.
\section{Discussion and conclusions}
\label{sec.06}
We calculated the bound-state spectrum of the Wick-Cutkosky model in
the ladder approximation both in light-front dynamics and in
instant-form dynamics. We used as a variational method an expansion
of the wave function in spline functions for the $x$ dependence and an
orthonormal set for the $p_\perp$ coordinate in LFD. Likewise, we used
a similar set for the radial wave functions in IFD. We checked the
absolute convergence of the spectra and also convinced ourselves that
the relative positions of the mass eigenvalues will very probably not
change at all when the number of basis functions would be increased above
what we have used.

As rotational invariance is broken in LFD by cutting off the Fock-space
expansion, we cannot expect the correct multiplet structure to appear
in the ladder approximation. Moreover, because the angular-momentum
operator is dynamical, we cannot immediately determine the angular
momentum quantum number $l$ of the states found in LFD.  Using
Terent'ev's transformation, which is  approximate in the bound-state
(off-energy-shell) case, we were able to characterize the bound states
by calculating their overlaps with the spherical harmonics
$Y_{lm}(\hat{p})$. Following this procedure, we could estimate the
degree of breaking of rotational invariance in LFD. Not surprisingly,
rotational symmetry is less violated for small values of the coupling
constant than for strong coupling.  

The well-known $O(4)$ symmetry of the WC-model is violated in the
ladder approximation in both LFD and IFD. If one computes the centers
of gravity of the states with the same $l$ and helicities $h = -l,
\dots, l$, one finds that the breaking of $O(4)$ symmetry in LFD and
IFD are quite similar.

In all cases we considered, the masses calculated in LFD are smaller
than in IFD. One might try to explain this fact by pointing out that
the Fock-space expansion converges more rapidly in LFD than in IFD, as
shown in Ref.~\cite{SBK98}, but one must keep in mind that in the
latter paper the Yukawa model was investigated, so one cannot be
absolutely certain of the validity of this argument for the WC-model.
Calculations done by Cooke and Miller~\cite{CM00} and Schoonderwoerd et
al.~\cite{SBK98} show that the breaking of rotational invariance in the
Yukawa model is partially restored by including the box diagram.

Although in the cases where the coupling constant is large, the bound
systems are certainly relativistic with binding energies of the same
order as the masses of the constituents, we find even then that the
binding energies follow the nonrelativistic rule $E_n = E_1/n^2$
remarkably well. This feature is illustrated in Table~\ref{NRcomp}. For
very weak coupling, $\alpha = 0.5$ and 1.0, we see a similar pattern
for the Salpeter equation, see Table~\ref{salpeter}, although in this
case the binding is much larger than in the LFD case.

\begin{table}[ht]
\caption{Comparison of the calculated binding energies (LFD and IFD) with
$E_{1}/n^{2}$ for the S-states at $\mu = 0$ in the case of $\alpha = 6$, 4 and 2.}
\begin{center}
\begin{tabular}{|c|c|c|c|c|c|c|}
\hline
\multicolumn{7}{|c|}{LFD} \\
\hline
& \multicolumn{2}{|c|}{$\alpha = 6$} & \multicolumn{2}{c|}{$\alpha = 4$} &
\multicolumn{2}{c|}{$\alpha = 2$} \\
\hline
& calc & $E_{1}/n^{2}$ & calc & $E_{1}/n^{2}$ & calc & $E_{1}/n^{2}$ \\
\hline
$E_{1}$ & 1.189 & 1.189 & 0.588 & 0.588 & 0.212 & 0.212 \\
$E_{2}$ & 0.256 & 0.297 & 0.144 & 0.147 & 0.055 & 0.053 \\
$E_{3}$ & 0.115 & 0.132 & 0.056 & 0.065 & 0.022 & 0.024 \\
\hline
\hline
\multicolumn{7}{|c|}{IFD} \\
\hline
& \multicolumn{2}{|c|}{$\alpha = 6$} & \multicolumn{2}{c|}{$\alpha = 4$} &
\multicolumn{2}{c|}{$\alpha = 2$} \\
\hline
& calc & $E_{1}/n^{2}$ & calc & $E_{1}/n^{2}$ & calc & $E_{1}/n^{2}$ \\
\hline
$E_{1}$ & 0.5069 & 0.5069 & 0.3334 & 0.3334 & 0.1526 & 0.1526 \\
$E_{2}$ & 0.1474 & 0.1274 & 0.0946 & 0.0834 & 0.0417 & 0.0382 \\
$E_{3}$ & 0.0696 & 0.0566 & 0.0442 & 0.0370 & 0.0192 & 0.0170 \\
\hline
\end{tabular}
\end{center}
%tab.06
\label{NRcomp}
\end{table}
\begin{table}
\caption{Bound state-masses calculated with a relativistic Salpeter
equation.}
\begin{center}
\begin{tabular}{|c|c|c|c|c|c|c|c|c|c|}
\hline
 & $\alpha = 0.5$ & $\alpha = 1$ & $\alpha = 2$ & $\alpha = 0.5$ & $\alpha = 1$ &
$\alpha = 2$ & $\alpha = 0.5$ & $\alpha = 1$ & $\alpha = 2$ \\
\hline
n & \multicolumn{3}{|c|}{$l = 0$} & \multicolumn{3}{|c|}{$l = 1$} & \multicolumn{3}{|c|}{$l = 2$} \\
\hline
1 & 1.9332 & 1.6825 & 0.4373 & 1.9842 & 1.9352 & 1.7092 & 1.9930 & 1.9719 & 1.8829 \\
2 & 1.9836 & 1.9251 & 1.6079 & 1.9930 & 1.9712 & 1.8710 & 1.9961 & 1.9842 & 1.9339 \\
3 & 1.9928 & 1.9683 & 1.8419 & 1.9961 & 1.9839 & 1.9289 & 1.9975 & 1.9899 & 1.9578 \\
4 & 1.9960 & 1.9827 & 1.9171 & 1.9975 & 1.9897 & 1.9553 & 1.9983 & 1.9930 & 1.9709 \\
5 & 1.9975 & 1.9891 & 1.9495 & 1.9984 & 1.9929 & 1.9694 & 1.9992 & 1.9948 & 1.9787 \\
\hline
\end{tabular}
\end{center}
%tab.07
\label{salpeter}
\end{table}
The potential used is not bounded from below and shows a collapse of
the wave function above a critical value of the coupling constant, see
Ref.~\cite{IBB00}. In the literature a critical value of $\alpha \approx
1.27$ can be found~\cite{Ray94}.  Looking at Table~\ref{salpeter} we see
that the ground-state mass for $\alpha = 2$ has dropped enormously,
illustrating the collapse.

An analytical solution of the Wick equation in the weak-binding limit
is given by Feldman et al.~\cite{FFT73}. In Table~\ref{wickeqn} the
bound-state masses are given for two values of the coupling constant,
$\alpha = 0.5$ and $\alpha = 1$, calculated with the expression for the
binding energy given in Ref.~\cite{FFT73}.  For these two values of the
coupling constant we also calculated the bound-state masses in LFD and
IFD (see Table~\ref{wickeqn}) for $l = 0$ only.

\begin{table}[ht]
\caption{Bound state-masses for the Wick equation in the weak-binding limit
\cite{FFT73} and the nS masses calculated in LFD and IFD.}
\begin{center}
\begin{tabular}{|c|c|c|c|c|c|c|c|}
\hline
 & \multicolumn{3}{|c|}{$\alpha = 0.5$} && \multicolumn{3}{|c|}{$\alpha = 1$} \\
\hline
n & Wick & LFD & IFD && Wick & LFD & IFD \\
\hline
1 & 1.965 & 1.973 & 1.9760 && 1.750 & 1.922 & 1.9364 \\
2 & 1.991 & 1.999 & 1.9939 && 1.938 &       & 1.9831 \\
3 & 1.996 &       & 1.9996 && 1.972 &       & 1.9931 \\
\hline
\end{tabular}
\end{center}
%tab.08
\label{wickeqn}
\end{table}

In his paper Cutkosky~\cite{WC54} gives a plot of his numerical
results obtained in the WC-model. From his figure we read off masses
for different values of the coupling constant (row `WC' in
Table~\ref{various}). In a similar way we read off the masses from
Fig.~2 in Ref.~\cite{DD00}. Looking at this figure we see that only the
results by Ji are not underestimating the WC-model masses. In
Table~\ref{various} we give the masses which we read off from the
curve labelled `Ji'. For comparison we also give the masses of the 1S
state calculated in the LFD approach and those calculated with the
analytical expression given by Ref.~\cite{FFT73} (row `FFT' in
Table~\ref{various}). Here we should remark that the analytical
expression is only valid for small values of the coupling constant.

\begin{table}[ht]
\caption{Comparison of different calculations of the 1S state for various
values of the coupling constant $\alpha$.}
\begin{center}
\begin{tabular}{|c|c|c|c|c|c|}
\hline
 & $\alpha = 0.5$ & $\alpha = 1$ & $\alpha = 2$ & $\alpha = 4$ & $\alpha = 6$ \\
\hline
WC &  1.97 & 1.92 & 1.77 & 1.39 & 0.60 \\
LFD & 1.973 & 1.922 & 1.788 & 1.412 & 0.811 \\
Ji & 1.98 & 1.93 & 1.82 & 1.49 & 1.09 \\
FFT & 1.965 & 1.750 & -0.765 & -30.241 & -130.192 \\
\hline
\end{tabular}
\end{center}
\label{various}
\end{table}

From Table~\ref{various} we see that the masses calculated by Ji
overestimate the bound-state masses for the WC-model. We also see that
the analytical expression of~\cite{FFT73} is only valid for small
values of the coupling constant $\alpha$.

Besides the mass spectrum we have also calculated both the
two-dimensional and radial wavefunctions. Breaking of rotational
invariance is clearly seen in the LFD case.
\end{document}